\documentclass[conference]{IEEEtran}
\usepackage{graphicx}
\usepackage{caption}
\usepackage{tikz}
\usepackage{subcaption}
\usepackage{hyperref}
\usepackage{amssymb}
\usepackage{amsfonts}

\usepackage{amsmath}
\DeclareMathOperator*{\argmin}{argmin}

\def\BibTeX{{\rm B\kern-.05em{\sc i\kern-.025em b}\kern-.08em
    T\kern-.1667em\lower.7ex\hbox{E}\kern-.125emX}}
\begin{document}

\title{ Learning Student Interest Trajectory for MOOC Thread Recommendation}
\author{\IEEEauthorblockN{ Shalini Pandey}
\IEEEauthorblockA{\textit{University of Minnesota} \\
\textit{Twin Cities}\\
Minnesota, USA, 55455 \\
pande103@umn.edu}
\and
\IEEEauthorblockN{Andrew Lan}
\IEEEauthorblockA{\textit{University of Massachusetts Amherst} \\
\textit{Amherst, MA 01003}\\
andrewlan@cs.umass.edu}
\and
\IEEEauthorblockN{ George Karypis}
\IEEEauthorblockA{\textit{University of Minnesota} \\
\textit{Twin Cities}\\
Minnesota, USA, 55455 \\
karypis@umn.edu}
\and
\IEEEauthorblockN{Jaideep Srivastava}
\IEEEauthorblockA{\textit{University of Minnesota} \\
\textit{Twin Cities}\\
Minnesota, USA, 55455 \\
srivasta@umn.edu}
}

\maketitle
\begin{abstract}
In recent years, Massive Open Online Courses (MOOCs) have witnessed immense growth in popularity. Now, due to the recent Covid19 pandemic situation, it is important to push the limits of online education. Discussion forums are primary means of interaction among learners and instructors. However, with growing class size, students face the challenge of finding useful and informative discussion forums. This problem can be solved by matching the interest of students with thread contents. The fundamental challenge is that the student interests drift as they progress through the course, and forum contents evolve as students or instructors update them. In our paper, we propose to predict future interest trajectories of students. Our model consists of two key operations: 1) Update operation and 2) Projection operation. Update operation models the inter-dependency between the evolution of student and thread using coupled Recurrent Neural Networks when the student posts on the thread. The projection operation learns to estimate future embedding of students and threads. For students, the projection operation learns the drift in their interests caused by the change in the course topic they study. The projection operation for threads exploits how different posts induce varying interest levels in a student according to the thread structure. Extensive experimentation on three real-world MOOC datasets shows that our model significantly outperforms other baselines for thread recommendation.
\end{abstract}

 \begin{IEEEkeywords}
 Recommender Systems, Personalized learning, MOOCs
 \end{IEEEkeywords}

\section{Introduction}
The world has transitioned into a new phase of online learning in response to the recent Covid19 pandemic. Now more than ever, it has become paramount to push the limits of online learning in every manner to keep flourishing the education system.
Massive Open Online Courses (MOOCs) provide a platform to teach students all kinds of subjects or courses. MOOCs have attracted a large number of users from across the globe, and the platform has uniquely enabled students to complete course exercises at their own pace and that too in an independent manner. However, learning from MOOC comes with its own set of challenges. One of the unique challenges faced by online learning platforms is that the means of interaction between students and instructors are critically limited. Peer learning, i.e., learning from each other through discussion, is an important component of the learning procedure and has a positive impact on student learning ~\cite{}. On MOOCs, discussion forums facilitate peer learning where instructors and students can ask questions, discuss ideas, and provide help to other students. However, as class size grows, the number of forums per course increases rapidly. As a result, it becomes quite difficult for a student to filter through a vast and overwhelming number of open forums to find relevant threads. 
\par
To address the information overload problem discussed above, it is necessary to build a thread recommendation system that yields a personalized shortlist of threads based on the student interest. Furthermore, the thread recommendation system in MOOCs helps decrease the amount of time required for new questions to go unanswered by directing appropriate users there ~\cite{yang2014forum}.
Traditional recommendation models have been used for recommending threads using collaborative filtering ~\cite{abel2009recommendations} and adaptive matrix factorization ~\cite{yang2014forum}. However, certain characteristics of thread recommendation on MOOCs set them apart from traditional recommendation systems. Firstly, MOOC forums are frequently updated by students or the instructors, which diversifies the content of these forums. Simultaneously, learners' preferences over MOOC topics evolve as they progress through the course; yet traditional recommendation techniques assume that learner interests and thread properties are static ~\cite{wu2017recurrent}. To capture this dynamic nature of the MOOC thread recommendation, a sequential recommendation model based on context tree ~\cite{garcin2013personalized} was proposed in ~\cite{mi2017adaptive}.   However, the main issue with such sequential recommendation models is that the student interest representation is updated only when an action (a reply on a thread or a post) occurs. However, \emph{ a student interest keeps evolving even when it has not taken any action}.  As a result, existing models are not entirely able to capture the dynamic nature of MOOC thread recommendation.    \par
In this paper, we propose a Student Interest Trajectory based Recommendation (SITRec), which represents students and thread as embedding vectors.
The evolution of student(and thread) is captured by a sequence of learned embeddings, which represents a trajectory of a student interests(and thread properties). Two key operations are employed to learn this trajectory: \textit{update operation} and \textit{projection operation}. The update operation updates the embedding of a student and a thread whenever an action involving the two is observed. It employs two mutually-recursive Recurrent Neural Networks (RNNs). One of them updates the student embedding using the thread embedding, and the other updates the thread embedding using the student embedding. \par
 Furthermore, even in the absence of any action, we update the embedding of students and threads using 
 the projection operation. The projection operation consists of two components: student projection operation and thread projection operation. The student projection operation is designed based on two intuitions. Firstly, as more time elapses time since  student's last update, her embedding will get farther. Secondly, the course topic that a student is studying at a time is a good indicator of her interest, and course topics are sequenced as defined in the course structure.  Thus, incorporating the course topic as a context feature in projection operation is beneficial in learning her projected embedding.  The thread projection operation learns personalized thread embedding for each student.
  Intuitively, student interest in a thread further increases, by different factors, if another student posts on it after the student's post or provide explicit comments to the student's post  ~\cite{lan2018personalized}. As a result, thread projection operation projects thread embedding with respect to student embedding based on nature of posts made on it after the student's post.  \par
 To predict the next thread which the student will be interested in, our model predicts embedding of the next thread. The recommendations can be made via nearest-neighbor search centered at that predicted embedding. \par
Extensive experimentation on real-world datasets shows that SITRec significantly outperforms the existing thread recommendation and dynamic embedding methods on Mean Average Precision (MAP). We conduct a comprehensive ablation study to show the effect of key components and visualize the drift in student interest and how it can be leveraged to find the topic of interest for each student. Summary of our paper major contributions are:
\begin{itemize}
    \item We consider the problem of thread recommendation as a dynamic sequential recommendation where both student embeddings and the thread embeddings keep evolving. We model the inter-dependency between the evolution of student and thread using mutually-recursive RNNs.
    \item We propose to predict student interest at a future time and then extract the relevant threads. We propose to utilize the course topic that the student is studying and elapsed time to predict the future interest of the student.
    \item We propose to project thread embedding personalized for each student so that we can incorporate how the interest of a student in a thread changes with the nature of posts made on the thread.
    \item We performed extensive experimentation involving an ablation study and visualizing the drift in student interest to support our methodology.
 \end{itemize}


\vspace{-3mm}
\section{Related work}
\subsection{Co-evolutionary models}
  Joint modeling of users and items, where each interaction between an item and a user updates the state of both interacting user and item has been explored in recommendation systems. RNNs have been used for modeling the evolving features of items and users in  ~\cite{ dai2016deep, kumar2018learning}. These models, similar to ours, also update the state of users and items after they interact. However, a major difference between these models and our work is that we take into account the course structure to further enhance the performance of our model. Additionally, we project the threads' embeddings personalized to each user such that we can take into account the likelihood of user posting on a thread because of the nature of past posts on the thread.
  \vspace{-3mm}
 \subsection{MOOC thread recommendation}
 MOOC has generated a huge amount of data attracting machine learning and data scientist for research. Here, we discuss all the research done for the recommendation of MOOC discussion forums. The works in  ~\cite{ramesh2014understanding} use unsupervised topic models with sets of expert-specified course keywords for capturing the category of forum posts. They, then, use topic assignments and sentiment to predict student course completion. Another work ~\cite{atapattu2016framework}  analyzed the content of the MOOC forum using topic modeling techniques to automatically generate labels for each thread. These labels can guide students in selecting interesting threads for themselves. Work in  ~\cite{inproceedings} couples social network analysis and association rule mining for thread recommendation; while their  approach considers social interactions among
learners, they ignore the content and timing of posts.
The adaptive matrix factorization based method  ~\cite{yang2014forum} groups learners  according to their posting behavior. It also studies the effect of window size, i.e., recommending only threads with posts in a recent time window. The work in~\cite{abel2009recommendations} uses a rule-based recommendation technique for providing personalized recommendations to individuals. However, these models do not capture the evolving features of users and threads. The point-process based method (PPS) proposed in ~\cite{lan2018personalized} models the probability that a learner makes a post in a thread at a particular time. This probability is computed based on interest level of the learner on the topic of thread,  timescale of the thread topic,  timing of the previous posts in the thread, and nature of the earlier posts regarding the learner. However, the user interest on a topic does not remain static across time. \par
As for modeling temporal dynamics, the work in ~\cite{brinton2014learning} proposed a method that classifies threads into different categories (e.g., general, technical, social) and ranks thread relevance for learners over time. However, it does not make personalized recommendations since it does not consider learners individually. Another work ~\cite{mi2017adaptive} leverages context trees that are used in a sequential recommendation system for providing adaptive recommendations.
MOOC forum recommendation differs from typical sequential recommendation problems because in MOOC forums, both student's interests and threads revolve around the course topics. The course structure is an additional source of information for predicting student interest and expertise. \par

To further facilitate personalization of online education, progress has been made in Knowledge Tracing ~\cite{pandey2019self, pandey2020rkt}, exercise recommendation ~\cite{huang2019exploring}, and Knowledge concept recommedation ~\cite{gong2020attentional} among others.
\begin{table}[]

\caption{Notations}
\label{notations}
\begin{tabular}{ll}
\hline
Notations & Description\\
\hline

$\boldsymbol{u}(t), \boldsymbol{p}(t) $                                             & Dynamic embedding of student $u$ and
thread $p$ \\ & at time $t$ \\

 $\boldsymbol{u}(t^-), \boldsymbol{p}(t^-) $                       & Dynamic embedding of student $u$ and thread p\\
 & right before time $t$  \\
$\bar{\boldsymbol{u}} , \bar{\boldsymbol{p}}  $                     & Static embedding of student $u$ and   thread $p$                                            \\

$\hat{\boldsymbol{u}}\boldsymbol{(t)}, \hat{\boldsymbol{p}}\boldsymbol{(t)}$ & Projected embedding of user $u$ and thread $p$ \\ &
at time $t$\\
$\tilde{\boldsymbol{q}}\boldsymbol{(t)}$ & Predicted embedding of post at time $t$\\

$\theta_t^{u,p}$ & Topic distribution of post made at   time $t$ \\

 $\Theta_i$ & Topic distribution of $i$th course topic taught in\\& the course.\\
 $\mathcal{P}_u(t)$ & Set of threads in which u has posted till time $t$\\
 $\mathcal{O}$& Set of posts in the training dataset. \\
 $t_{u,p} $ & The last time user $u$ posted on thread $p$.\\
 $\textbf{w}_t^{u,p}$ & The term-frequency vector of the post made by \\ &  user $u$ on thread $p$ at time $t$.\\

\hline
\end{tabular}

\end{table}
\captionsetup[subfigure]{font=small,skip=0pt}
\captionsetup[figure]{font=small,skip=0pt}
\begin{figure}
    \centering
    \includegraphics[width=0.5\textwidth]{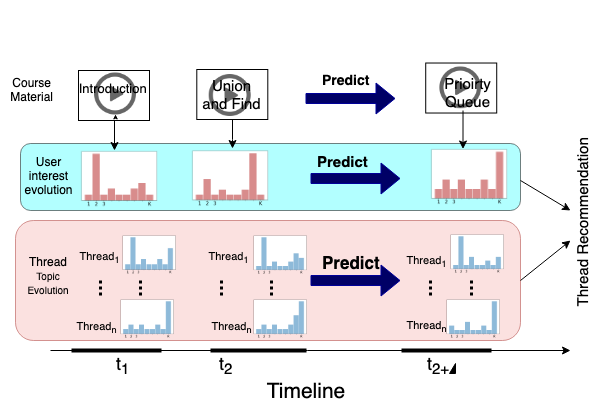}
    \caption{Temporal evolution of student interest and thread topics. The orange bar chart shows the interest level of students on the topics $[1, \ldots, K]$. The blue bar chart shows the  probability of thread's content to belong to topics $[1, \ldots, K]$}
    \label{Evolution}
\end{figure}

\vspace{-3mm}
\section{Proposed Method}
\textit{Problem statement:} In the setting of thread recommendation, we are given $m$ students, $n$ threads, and $N$ posts. Each post can be represented as a tuple,
$(u, p, t, \textbf{w}_t^{u,p})$, where $\textbf{w}_t^{u,p}$ denotes the term-frequency vector of the post made by student $u$, on thread $p$ at time $t$. The notations used in this paper are described in Table ~\ref{notations}. The problem of thread recommendation can be defined as: For each student, $u$, find the most relevant threads that she will be interested in.  As shown in Figure ~\ref{Evolution} the thread content as well as student interest keeps evolving with time. The interest of the student is further affected by the course topic she
is studying. To perform the recommendation task, it is important to predict the future interest of students and properties of threads. \\
\textit{Overview:}
Our model, SITRec learns embeddings to represent student interests and thread properties. Overall, SITRec comprises of two major operations: update operation and projection operation. The update operation uses two mutually-recursive RNNs to update the embedding of student and thread after the student posts on the thread. To predict student embedding at a future time, the student projection operation leverages the course structure and the elapsed time since last update of student embedding. Lastly, our model also generates a student personalized projected thread embedding which takes into account the idea that a student is more likely to post on the thread she is already associated with. This behavior replicates the notification setting for MOOCs.


\subsection{Text Representation}
The text of each post can be represented as a distribution over few topics because the posts in MOOCs are centred around the topics associated with the course. For this reason, we use topic modeling technique to extract text feature from the post. For extracting the features, we build a dictionary of item vocabularies after filtering the stop words
and removing words that occur fewer than $10$ times. The content of each post text can be represented as a vector, $\textbf{w}_t^{u,p} = {[w_{t1}^{u,p}, w_{t2}^{u,p},\ldots, w_{t_W}^{u,p}]}$, where  $W$ is the total number of words in the vocabulary and $w_{tj}^{u,p}$ represents the frequency of word $j$ in the post. The topic distribution vector $\theta_{t}^{u,p}$  is used to represent the post by student  $u$ in thread $p$ at time $t$ and computed using the Latent Dirichlet Allocation (LDA) model~\cite{blei2003latent}. \par

To find the features associated with the course topic taught in $i$th week of the course, $\Theta_i$, we use LDA to extract the topic distribution of the description of all course materials taught in the $i$th week. This description of course materials is extracted from the synopsis of course obtained from their respective website \footnote{https://www.coursera.org/}.
\captionsetup[subfigure]{font=small,skip=0pt}
\captionsetup[figure]{font=small,skip=0pt}
\begin{figure}[t]
       \includegraphics[width=0.5\textwidth]{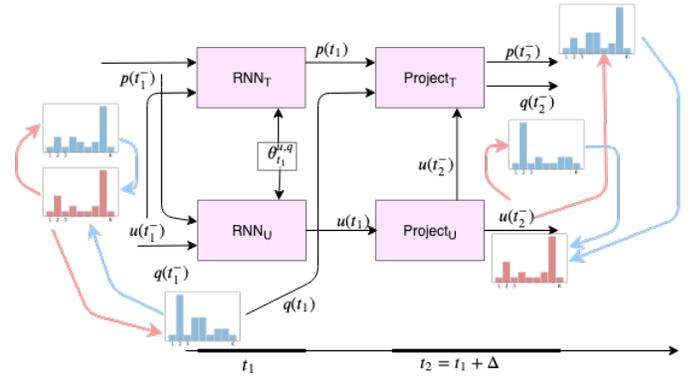}
     \caption{The SITRec model: Model illustation for student $u$ (orange) and threads $p$ and $q$ (blue).  Student features and thread features influence each other and co-evolve with time.  At time $t_1$, student $u$ posts on thread $p$, the dynamic embeddings of both $u$ and
  $p$ are updated with RNN\textsubscript{U} and RNN\textsubscript{T},  respectively.
 The projection operation Project\textsubscript{U} and Project\textsubscript{T} predicts the student and thread embedding , respectively at a future time ($t_1+\Delta$).}
     \label{architecture}

   \end{figure}
  \vspace{-4mm}
\subsection{Embedding layer}
 We assign each student and thread two embeddings: a static and a dynamic embedding. The static embedding for student $\bar{\boldsymbol{u}} \in \mathbb{R}^m$, encodes  the general interest or expertise (which represent the likelihood of student to post on  a thread) of students, while that for threads, $\bar{\boldsymbol{p}} \in \mathbb{R}^n$, represents the main topic focussed in the thread. They are obtained using one-hot vectors as inputs, as described in ~\cite{zhu2017next}. The dynamic embedding of student, $\boldsymbol{u}(t) \in \mathbb{R}^d$ 
 changes with time and is used to capture the evolving student interest. Similarly, the discussion in threads sometimes deviate as new posts and comments are added. In order to model this dynamic nature, we employ dynamic embedding for each thread, $\boldsymbol{p}(t)\in \mathbb{R}^d$. \par
\subsection{Update operation}
Whenever a student posts on a thread, both thread embedding and student embedding gets updated. This update is modeled by two mutually-recursive Recurrent Neural Networks (RNNs).
  The hidden states of the RNN\textsubscript{U} and the RNN\textsubscript{T} represent the student and thread embeddings, respectively.
The two RNNs are coupled together because thread embedding affects the student embedding and student embedding affects that of thread. As shown in Figure ~\ref{architecture}, when student $u$ posts on thread $p$, RNN\textsubscript{U} updates the embedding $\boldsymbol{u}(t)$ by using the embedding $\boldsymbol{p}(t^-)$ of thread $p$ right before time $t$ and text representation of the post $\theta_t^{u,p}$ as inputs. Similarly, RNN\textsubscript{T} updates embedding $\boldsymbol{p}(t)$ by using the embedding $\boldsymbol{u}(t^-)$ of student $u$ right before time $t$ and text representation of the post $\theta_t^{u,p}$ as inputs.  More formally,
\begin{equation*}
    \boldsymbol{u}(t) = \sigma (\boldsymbol{W}^{u}[\boldsymbol{u}(t^-), \boldsymbol{p}(t^-), \theta_t^{u,p} ,  \Delta_u]),
\end{equation*}
\begin{equation*}
    \boldsymbol{p}(t) = \sigma(\boldsymbol{W}^p[\boldsymbol{p}(t^-), \boldsymbol{u}(t^-), \theta_t^{u,p},\Delta_p]),
\end{equation*}
where $\Delta_u$ denotes the time since $u$'s previous post on any thread and $\Delta_p$ is the time since last post on thread $p$, $\theta_t^{u,p}$ is the text feature vector of the post. The matrices $\boldsymbol{W}^u, \boldsymbol{W}^p \in \mathbb{R}^{(2d+F+1)\times d}$ are the parameters of RNN and $F$ is the number of features associated with the post.
\captionsetup[subfigure]{font=small,skip=0pt}
\captionsetup[figure]{font=small,skip=0pt}
\begin{figure}[!ht]
       \includegraphics[width=0.5\textwidth]{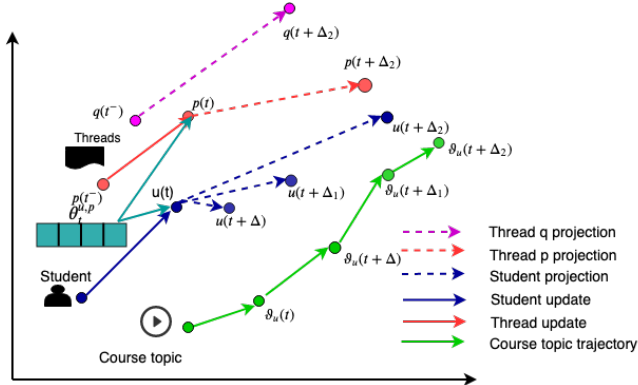}
     \caption{Projection Operation : This figure shows the key idea behind projection operation. At time $t$, student $u$ posts in thread $p$ with post features $\theta_t^{u,p}$. The projected embedding of student $u$ is shown for
different elapsed times $\Delta < \Delta_1 < \Delta_2$. The course topics $\vartheta_u$ represents the topics $u$ is studying at different times.  The embeddings of the two threads, $p$ and $q$ are also shown. After elapsed time $\Delta_2$ thread $p$'s embedding is projected closer to $u$'s embedding while thread $q$ on which $u$ did not post in the past is projected farther from $u$'s embedding.  }
     \label{projection}
   \vspace{-4.5mm}
   \end{figure}

\subsection{Projection Operation}
The projection operation predicts the future trajectory of  student  interests based on course structure and student personalized embeddings of threads based on the nature of new posts made on the thread.
\subsubsection{Student Projection}

In this section, we will describe how we obtain the future embedding trajectory of a student. The motivation behind the student projection operation is two-folds: 1) as time elapses student interest drifts farther from the original, 2) the course topic a student is interested in is an important factor in deciding her future 
interest.  As shown in Figure ~\ref{projection}, a student $u$ posts at time $t$ and the RNN layer outputs her interest embedding $\boldsymbol{u}(t)$. After a short duration $\Delta_1$ since $t$, the student's projected
embedding $\boldsymbol{u}(t + \Delta_1)$
 is close to her previously observed embedding
$\boldsymbol{u}(t)$. As more time $\Delta_2 > \Delta_1 > \Delta$ elapses, the projected embedding
drifts farther from $\boldsymbol{u}(t)$ and the course topic embedding ,$\vartheta_u(t)$, helps in guiding the evolution of projected embedding of the student.\par

The first step in projecting a student embedding is to determine the topic, she is interested in, which is determined as,
   $ \vartheta_u(t) = i | \Theta_i = \argmin_\Theta ||\Theta-\theta_t^{u,p}||_2$, 
   where $\Theta_i $ is the topic distribution of $i$th week course content and $\theta_t^{u,p}$ is the topic distribution of post made by student $u$ on thread $p$ at time $t$. Then, to predict the projected student embedding, we
   incorporate the context features:  current course topic embedding, $\vartheta_u(t)$ and the elapsed time since last update, $\Delta$ along with student current embedding $\boldsymbol{u}(t)$ as input. Since simply concatenating the context features and passing through linear layer has proved to be ineffective in modeling the interaction between the concatenated input features, we follow the procedure suggested in Latent Cross ~\cite{beutel2018latent}.  We describe how we obtain the feature-context vector below.\par
  To incorporate the context feature $f$, we first convert $f$ to a feature-context vector $\boldsymbol{w}_f \in  R^d$ using a linear
layer  $\boldsymbol{w}_f = \boldsymbol{W}_ff$. The weights of the linear layer, $\boldsymbol{W}_f$ is initialized by a $0$-mean Gaussian. 
We represent the time-context vector as $\boldsymbol{w}_\Delta$ and the course topic-context vector as $\boldsymbol{w}_\vartheta$. The projected embedding is then obtained as element-wise product of the context vector and the previous embedding as,
  \begin{equation}
      \hat{\boldsymbol{u}}(t+\Delta)= (\boldsymbol{1}+\boldsymbol{w}_\Delta+\boldsymbol{w}_\vartheta ) * \boldsymbol{u}(t)
  \end{equation}
\subsubsection{Thread Projection}
Thread projection layer projects thread embedding personalized to each student based on the nature of posts made on the thread. It is essentially important to capture the temporal dynamics of threads. Intuitively, a student is likely to be interested in a thread if another student posts on the thread which she is already associated with. The level of interest further increases if another student  comments on the student's post. This also reflects the notification setting for discussion forum, where student gets notified whenever any posts/comments are made on threads that the student has interacted with. Motivated by this, we develop a thread projection layer which learns a student-personalized thread embedding such that the thread embedding is projected closer to student embedding based on nature of posts/comments made on the thread after the student's last interaction.  \par
%
The projected thread embedding with respect to student $u$ is obtained as,
\begin{equation}
    \boldsymbol{\hat{p}}_u(t+\Delta) = \frac{\zeta_{u,p}(t+\Delta)}{1+\zeta_{u,p}(t+\Delta)}\boldsymbol{u}(t)+ \frac{1}{1+\zeta_{u,p}(t+\Delta)}\boldsymbol{p}(t),
\end{equation}
where $\zeta$-factor, $\zeta_{u,p} (t+\Delta)$ defines how much closer the projected thread embedding is to the student embedding. The higher the value of $\zeta$-factor, the closer is the projected thread embedding to the student embedding.  Naturally $\zeta$-factor should have different terms for posts on the thread and comments on student's post as they induce different level of excitement among students ~\cite{lan2018personalized}. This excitement also fades as the time elapses owing to the ageing of the threads. As a result we define $\zeta$-factor as:
\begin{align}
 \begin{split}
    \zeta_{u,p}(t+\Delta) = \boldsymbol{1}_{\mathcal{P}_u}
   ( \sum_{t_{u,p}<t_p<t+\Delta} e^{-\alpha (t_p-t_{u,p}) } \\+
    \sum_{t_{u,p}<t_r<t+\Delta} e^{-\beta (t_r-t_{u,p})}),
\end{split}
\end{align}
\noindent where  $\boldsymbol{1}_{\mathcal{P}_u}$ is $1$ if $u$ posted in $p$ , otherwise $0$, $t_{u,p}$ is the last time user $u$ posted on $p $, $\alpha$ and $\beta$ are the scalar weights given to the excitement level induced by a new post  and replies on the student's posts on the thread $p$, $t_p$ and $t_r$ are the timestamps of posts made on the thread $p$  and the timestamps of the explicit replies made on the student's post on $p$, respectively. \par
\vspace{-2mm}
\subsection{Recommendation}
Similar to JODIE model ~\cite{kumar2018learning}, we predict the embedding of the next thread that will interest the student. We make this prediction using the projected student embedding
$\hat{\boldsymbol{u}}(t + \Delta)$ and the embedding of thread $\boldsymbol{p}(t)$ of thread $p$ (the thread on which $u$ last posted on). The reason we
include $\boldsymbol{p}(t)$ is that students often interact with the same item
consecutively and including the item embedding helps to
ease the prediction. The prediction is made using a  linear layer as
follows:
\begin{equation*}
    \tilde{\boldsymbol{q}}(t + \Delta) = \boldsymbol{W}[ \boldsymbol{\hat{u}}(t + \Delta) ,  \boldsymbol{\bar{u}} ,  \boldsymbol{p}(t ), \boldsymbol{\bar{p}}] + \boldsymbol{B}  ,
\end{equation*}

\noindent where $\boldsymbol{W} \in \mathbb{R}^{(m+n+2d)\times (n+d)}$ is the weight matrix and $\boldsymbol{B} \in \mathbb{R}^{n+d}$ is the bias vector in the linear layer. \par
Having generated the predicted thread embedding at time $t+\Delta$, we find the candidate threads for recommendation using nearest-neighbor search which are closest to the predicted thread embedding.
\vspace{-2mm}
\subsection{Network Training}
We train our model to minimize the Euclidean distance between the predicted thread embedding and the ground truth thread embedding everytime a student posts on a thread. We calculate the total loss as,
\begin{align*}
 \begin{split}
     Loss = {}&\sum_{u,p,t,\boldsymbol{w}_t^{u,i}\in \mathcal{O}} ||\tilde{\boldsymbol{q}}(t)-[\boldsymbol{\bar{p}}, \boldsymbol{\hat{p}}_u(t)]||_2 + \\\lambda_U ||\boldsymbol{u}(t)-\boldsymbol{u}(t^-)||_2
     & + \lambda_T ||\boldsymbol{p}(t) - \boldsymbol{p}(t^-)||_2,
 \end{split}
\end{align*}
\noindent where $\mathcal{O}$ is  set of posts in  training sample, $\lambda_U$ and $\lambda_T$ are regularization parameters for temporal smoothness of student and thread embeddings, respectively. The complete parameter space in our training models is
$\Omega_\text{update} = \{ \boldsymbol{W}^u, \boldsymbol{W}^p\}, \Omega_\text{project}= \{\boldsymbol{W}_\Delta, \boldsymbol{W}_\vartheta, \alpha, \beta\}, \Omega_\text{Rec} = \{ \boldsymbol{W},\boldsymbol{B} \} , \Omega_{reg} = \{ \lambda_U , \lambda_T\}$.

 \captionsetup[subfigure]{font=small,skip=0pt}
\captionsetup[figure]{font=small,skip=0pt}
\begin{figure*}

\begin{minipage}[b]{0.35\textwidth}
\includegraphics[width=.9\linewidth]{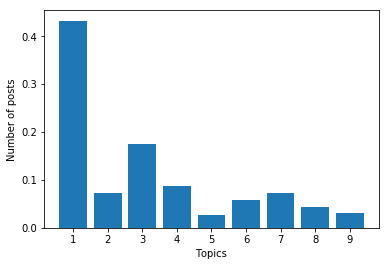}\\
\subcaption{Algo}
\end{minipage}%
\begin{minipage}[b]{0.35\linewidth}
\includegraphics[width=.9\linewidth]{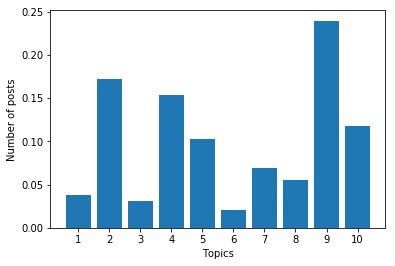}\\
\subcaption{ML}
\end{minipage}%
\begin{minipage}[b]{0.35\linewidth}
\includegraphics[width=.9\linewidth]{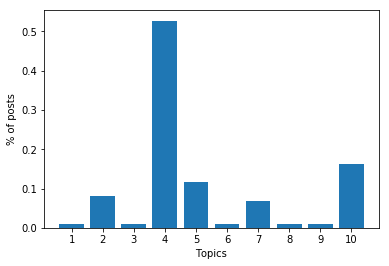}\\
\subcaption{comp}
\end{minipage}%
\caption{ Dataset statistics in terms of posts per topic.}
\label{data}
\end{figure*}
\section{Experimental Settings}
To comprehensively evaluate the performance of our proposed SITRec model, we design different strategies to evaluate the effectiveness of the model.
\begin{table}[]
\centering
\caption{Dataset Statisitics}
\label{Dataset}

\begin{tabular}{lrrrrr}
\hline
\textbf{Dataset}  & \textbf{Threads} &    \textbf{Posts} & \textbf{Learners} & \textbf{Weeks}   \\
\hline
\textbf{ml} & 5310 & 40050 & 6004    &15                                         \\
\textbf{algo}    & 1323 &    9274    & 1833        & 9\\
\textbf{comp} & 4860 &17562 &3060    & 14          \\
\hline
\end{tabular}
\vspace{-1em}
\end{table}
\vspace{-2mm}
\subsection{Dataset}

We use three real-world datasets to evaluate the performance of our model. These datasets are obtained from Coursera course offering for three courses, namely, Machine Learning (ml), Algorithms, Part I (algo), and English Composition I (comp), in 2012. Table ~\ref{Dataset} gives details on these datasets.
 These datasets, in addition to varying in the size of users and density of interaction, also comprises of different user behavior in terms of posts per topic. As shown in Figure ~\ref{data}, ml has the most diversified posts, pertaining to different topics, while algo and comp have most posts related to one topic.
\vspace{-2mm}
\subsection{Comparison Approaches}
We compare our model with the following approaches: \par

\begin{itemize}

    \item \textbf{Popularity-based (POP)}: This is a simple baseline that ranks threads from most to least popular according to their popularity.

\item \textbf{Recency-based (REC)}: This is also a simple method that ranks threads from oldest to newest based on the time the most recent post was made on the thread.

\item \textbf{Personalized Recency-based (USER-REC)}: This method ranks threads from oldest to the newest based on the time the user interacted with the thread.

\item \textbf{Adaptive Matrix Factorization (AMF) ~\cite{yang2014forum}}: This is an Adaptive Matrix Factorization based method which finds similar users and recommends those threads to a user which similar users have posted on.

\item \textbf{Point Process based (PPS) ~\cite{lan2018personalized}}: This is a Point Process based method which calculates the probability that a user will post on a thread. It uses a heuristic that a post on a thread and an explicit reply on a user's post increases the likelihood of participation of the user on the thread in different manner.
\item \textbf{Deep Coevolutionary (DeepCo-evolve)~\cite{dai2016deep}}: A co-evolutionary model that updates user and item embeddings when a user interacts with an item using RNN. To predict whether user will interact with item it employs  point process technique where the probability of the interaction  decays with time.
\item \textbf{JODIE ~\cite{kumar2018learning}}: JODIE is state-of-the-art model for predicting a user's interaction with item. It is also co-evolutionary model that projects user embedding using temporal attention layer after some elapsed time $\Delta$  since user's previous interaction.
\end{itemize}

  \captionsetup[table]{font=small,skip=0pt}
\begin{table}[]
\begin{center}
\caption{ Performance comparison on three datasets for all methods  in terms of Mean Average Precision (MAP) $@5$. The best and the second best results are highlighted by
\textbf{boldface} and \underline{underlined } respectively. Gain$\%$ denotes the performance improvement of SITRec over the best baseline.}
    \begin{tabular}{|c|r|r|r|}
    \hline
\textbf{Methods} &        \textbf{Algo}              &      \textbf{ML}                &        \textbf{Comp}              \\
\hline
 POP&  0.102 &  0.005                 &    0.001                                     \\
 \hline
 REC&  0.020 &      0.090             &         0.066                                 \\
 \hline
 USER-REC  &    0.338   &0.150             &               0.221                        \\
 \hline
 AMF ~\cite{yang2014forum}& 0.091 &   0.005                 &  0.253                                        \\
 \hline
 PPS ~\cite{lan2018personalized}& 0.362  &     0.152           &  \underline{0.332}                              \\
 \hline
 DeepCo-evolve ~\cite{dai2016deep}&0.112 &0.088 & 0.162 \\
 \hline
 JODIE ~\cite{kumar2018learning} & \underline{0.397}&\underline{0.253}  & 0.212\\
 \hline
 SITRec  & \textbf{0.561}&\textbf{0.400} & \textbf{0.393} \\
 \hline
 Gain$\%$ & 41.310 & 58.102& 18.373\\

\hline
\end{tabular}

\label{methods}
\end{center}

\end{table}

\subsubsection{Metrics}
We evaluate forum recommendation using the standard ranking metric Mean Average Precision ($MAP@N$).  In our experiments,  we set $N = 5$.
\begin{equation}
    AP_u@N =\frac{ \sum_{n=1}^N P_u@n\times post(n)}{min{|R_u,N|}},
\end{equation}
\noindent where $R_u$ is the set of threads student $u$ posted on during
the test time interval and $post(n)$ is a binary
function that describes whether the user has posted in the $n$th thread. $P_u@n$ denotes the precision at $n$.  Finally, $MAP$  is obtained by averaging the $AP$ values of all the users.

\subsection{Evaluation Methodology}

\subsubsection{Model Training and Parameter Selection}

We perform a series of pre-processing on the text of posts. For preparing the feature associated with each post we process the text by i) removing url links, punctuations and words that contain digits, ii) convert all words to their respective base forms, iii) remove stopwords and (iv) remove words that appear fewer than 10 times. Then, we obtained a bag-of-words representation of each text. The process used for obtaining features associated with each post from bag-of-words representation is explained in Section 3.1.  The number of topics used in LDA algorithm is same as the number of topics in the course as extracted from the course syllabus because we assume that forums are centered around the topics of course content. We also run LDA on the course syllabus obtained from the course website. \par

For all the datasets, we tried the embedding dimensions from $[5,10,15,20,25]$  and chose the value that gave the best performance. The values of $\alpha$ and $\beta$ required in thread projection were selected from $[0.0005,0.001, 0.005,0.01,0.05,0.1,0.5, 1.0]$.
We found that $\alpha = 0.5$ and $\beta = 0.001$ gives the best performance for algo dataset, $\alpha$ = $0.5$ , $\beta$ = $0.005$ gives the best performance for ml dataset and $\alpha$ =$0.5$, $\beta=0.1$ gives the best performance for comp dataset.
We used learning rate of $0.001$ and  t-batch algorithm ~\cite{kumar2018learning} for creating the batches in our experiments. \footnote{Code available at \url{https://github.com/shalini1194/MOOC_forum_recommendation}}

\begin{figure*}
   \centering
\includegraphics[width= 0.9\textwidth]{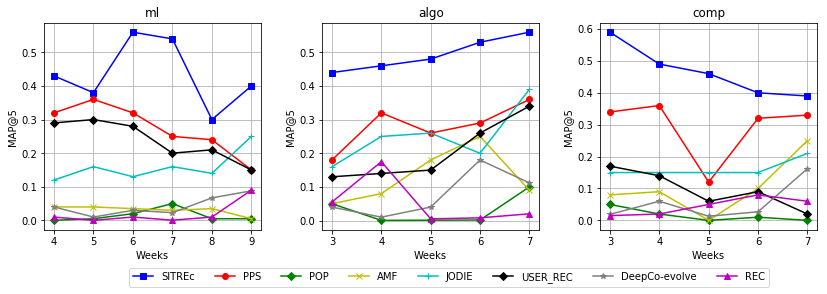}



\caption{Plot of recommendation performance over different lengths of the training time window $T1$ on all datasets. Our model, SITRec  significantly outperforms every baseline.}
 \label{performance1}

 \end{figure*}

\section{Results and Discussion}

\subsection{Performance Evaluation }
Table ~\ref{methods} shows the  the recommendation performance of our model and the baselines over for all the dataset when the training time $T_1$ is set to  $W-1$ week, where $W$ is the duration in which forums are active and testing time interval ($T_2-T_1$) is one day after. The value of $W$ is $10, 8 , 8$ for ml, algo, and comp, respectively. Since learners drop out from the course with time leading to reduction in forum activities, these values of $W$ is less than those  mentioned in Table ~\ref{Dataset}. \par
As seen in the table ~\ref{methods}, our proposed SITRec significantly outperforms existing methods in all the datasets. Among the simple  baselines (POP, REC, USER-REC),  USER-REC performs better than the rest. This confirms that users tend to post comments on threads they are already associated with.  USER-REC performs better than AMF because AMF does not take into account the posts that the user has already posted on. Since repetitive behavior of users is an important signal for making prediction of next thread and AMF fails to take that into consideration, it is outperformed by USER-REC and PPS methods. Among the co-evolutionary models proposed in the literature, JODIE significantly outperforms Deep Co-evolve which is in agreement to ~\cite{kumar2018learning}. Since JODIE takes into consideration the last thread on which user posted to predict the embedding of the next thread, it performs better than DeepCo-evolve Finally, SITRec outperforms all the baselines. There is no clear winner among the baselines: JODIE performs better than PPS on algo and ml dataset while PPS performs better than JODIE on comp dataset. This could be because in comp being English Composition dataset, the discussions in each thread is longer, leading to more activity notifications and students tending to reply on same thread, while in engineering courses like ml and algo learners are expected directly answer each other's questions than holding long discussion ~\cite{lan2018personalized}. \par
The fact that SITRec outperforms the JODIE
baseline confirms both our hypothesis regarding MOOC
forums. First, it is important to consider how the user's
interest evolves (by taking into account the course topic
that the student is studying). Second, user's interest in
a thread increases if she has already posted in that thread
and if someone replies on his post. The fact that SITRec
outperforms other baselines which do not consider the evolving nature of user interest and thread's properties emphasizes the benefit of the co-evolutionary RNNs in capturing the dynamic nature involved in thread activities.   \par

\begin{figure}[t]
\centering
  \includegraphics[width=0.35\textwidth]{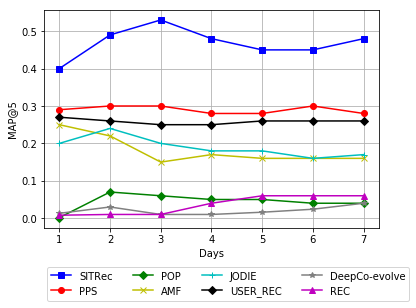}

\caption{Recommendation performance for algo dataset
by varying testing window length.
}
\label{performance2}

\end{figure}

\subsubsection{Robustness towards proportion of data}
In this experiment, we validate the robustness of SITRec by varying
the data taken as training set and test set and comparing the performance of the algorithm with other baseline methods.
In the first setting, we hold  the testing interval fixed to one day and vary the training data size from $1$ week to  $W-1$ weeks, where $W$ is the duration in which forums are active and testing time is one day after. Figure ~\ref{performance1} shows the performance of the various methods in this setting. Overall, we see that our model significantly outperforms the baselines in each case, achieving $7\%$ to $190\%$ improvement over the strongest
baseline. Another interesting observation is that even
when the training data is of small interval, (i.e., when training data consisted of only few weeks) SITRec gives good performance compared to other models.  \par
In the second setting, we hold the length of the training interval
fixed at $W-2$ weeks to allow sufficient number of posts in the test week and vary the length of the testing interval from $1$ day to $7 $ days.
Figure ~\ref{performance2} shows the recommendation performance over different lengths of the testing time window $\Delta T$ for the algo dataset. Our model, SITRec outperforms the baseline methods for all the values of $\Delta T$. Even when the length of test set interval increases the performance of our model does not degrade, in fact improves in some cases. This can be explained by the intuition that every action taken by a student improves the learnt embedding of student interest. Thus, our model is robust towards the length of testing interval and is able to model student behavior over long period of time as well.  Since the performance on other datasets was similar, we omitted the chart of other datasets.
 \vspace{-2mm}
\subsection{Ablation Study }
In order to verify the effectiveness of the modification we introduced in this paper, we run an ablation study to check the importance of each individual component. The results are provided in Table ~\ref{Ablation}. The variants of our models are:

\captionsetup[table]{font=small,skip=0pt}
\begin{table}[t]
\centering
\caption{Comparing variants of the proposed model.
Best results are indicated in \textbf{bold}}
\begin{tabular}{|l|r|r|r|}

\hline

\textbf{Dataset}                 & \textbf{algo}          & \textbf{ml}            & \textbf{comp}          \\ \hline
SITRec-Dynamic Student      & 0.37          & 0.38          & 0.29          \\ \hline
SITRec-Dynamic Thread    & 0.44          & 0.35          & 0.23          \\ \hline
SITRec-Student Projection   & 0.53          & 0.37          & 0.30           \\ \hline
SITRec-Thread Projection & 0.16          & 0.21          & 0.05          \\ \hline
SITRec-Text Features              & 0.46          & 0.34          & 0.27          \\ \hline
SITRec                   & \textbf{0.56} & \textbf{0.40} & \textbf{0.39} \\ \hline

\end{tabular}
\label{Comparison}

\label{Ablation}

\end{table}
\vspace{-1mm}
\begin{itemize}

    \item \textbf{SITRec-Dynamic Student}: We remove the dynamic embedding of student in this variant of SITRec model.

    \item \textbf{SITRec-Dynamic Thread}: We remove the dynamic embedding of thread in this variant of SITRec model.

     \item \textbf{SITRec-Student Projection}: In this variant,  we do not predict future student embedding and the embeddings are only updated when student makes a post on a thread.

    \item \textbf{SITRec-Thread Projection}: In this variant of SITRec model, we do not project a thread embedding specific to each student. We  use the embedding of thread obtained right after the update operation.

    \item \textbf{SITRec-Text Features}: In this variant of SITRec, we remove the text feature input to  RNN\textsubscript{U} and RNN\textsubscript{T} models.

    \end{itemize}

The results are obtained by taking $W-1$ weeks as training interval and $1$ day as testing interval for each dataset.
To demonstrate that student embeddings and thread embeddings change with time, we compute the performance of model with only static student and thread embedding. The decrease in performance implies that it is important to incorporate the dynamic nature of students and threads to deliver the best results.
 Removal of the student projection operation is shown to reduce the performance of model to some extent, however, removal of thread projection causes a drastic reduction in performance of the model. This suggests that on MOOC forums students tend to post on the threads they have already visited before. This factor plays an important role in deciding the thread to recommend when multiple threads on same topic exist. Without thread projection layer, even if SITRec predicts correct topic of interest for student, it fails to identify particular thread to recommend. The drop in performance for ml is less pronounced than that in algo and comp. To investigate further, we plot how diverse threads are in these datasets and find that in ml the distribution of threads is fairer while that in algo and comp is skewed towards a particular topic. As a result, identifying the thread after determining the topic of interest is easier for ml dataset compared to algo and comp.
 At last, to investigate the effectiveness of textual features of posts and comments in a thread, SITRec-Text feature is introduced where no textual features are fed to the two RNNs in the update operation. We find decrease in performance of the model suggesting that textual features help in enhancing the model performance.
 \vspace{-2mm}
 \section{Conclusion and Future Work}

In this paper, we proposed a student interest trajectory based solution to MOOC thread recommendation problem. Our method, SITRec models the dynamic nature of student interest and thread contents. It also leverages the course topic structure and how student interest towards a thread changes when posts are made on a thread that the student has already interacted with. This captures the temporal dynamics of posting behavior of students in online forums. We demonstrate the superiority of the
performance of our model compared to other competing approaches on three real-world datasets.\par
As part of future work, we plan to incorporate the social structure i.e., how post made by a friend influences a student's interest in a thread. Along with that, we plan to explore the the other applications where student interest trajectory prediction can be useful such as external content recommendation.

\bibliographystyle{ACM-Reference-Format}

\bibliography{reference.bib}

\end{document}